\documentclass[12pt]{article}
\usepackage{graphicx}

\usepackage{amssymb}
\usepackage{amsmath}
\usepackage{float}	
\usepackage{subfigure}
\usepackage{multirow}
\usepackage{float}
\restylefloat{table}

\newcommand\pubnumber{WSU--HEP--XXYY}
\newcommand\pubdate{\today}

\def\wayne{Department for Experimental Particle Physics \\ Jozef Stefan Institute, Jamova 39, Ljubljana, Slovenia}
\def\support{\footnote{on behalf of the Belle Collaboration.}}

\def\Title#1{\begin{center} {\Large #1 } \end{center}}
\def\Author#1{\begin{center}{ \sc #1} \end{center}}
\def\Address#1{\begin{center}{ \it #1} \end{center}}

\newcommand\pubblock{\rightline{\begin{tabular}{l} \pubnumber\\
         \pubdate  \end{tabular}}}
\newenvironment{Abstract}{\begin{quotation}  }{\end{quotation}}
\newenvironment{Presented}{\begin{quotation} \begin{center} 
             PRESENTED AT\end{center}\bigskip 
      \begin{center}\begin{large}}{\end{large}\end{center} \end{quotation}}
\def\Acknowledgements{\bigskip  \bigskip \begin{center} \begin{large}
             \bf ACKNOWLEDGEMENTS \end{large}\end{center}}

\def\beq{\begin{equation}}
\def\eeq#1{\label{#1}\end{equation}}
\def\eeqn{\end{equation}}

\def\beqa{\begin{eqnarray}}
\def\eeqa#1{\label{#1}\end{eqnarray}}
\def\eeqan{\end{eqnarray}}

\let\bar=\overbar

\def\Dslash{\not{\hbox{\kern-4pt $D$}}}
\def\dslash{\not{\hbox{\kern-2pt $\del$}}}

\def\msb{{\bar{\ssstyle M \kern -1pt S}}}

\begin{document}
\begin{titlepage}
\pubblock

\vfill
\Title{Search for CP violation in $D^0 \to K^+K^-,\pi^+\pi^-$ and $D^0 \to \pi^0\pi^0$}
\vfill
\Author{Tara Nanut\support}
\Address{\wayne}
\vfill
\begin{Abstract}
We report updated measurements of indirect and direct CP asymmetry in decays $D^0 \to K^+K^-,\pi^+\pi^-$ and a new measurement of direct and indirect CP asymmetry in decays $D^0 \to \pi^0\pi^0$, using the full data sample of the Belle experiment.
\end{Abstract}
\vfill
\begin{Presented}
The 7th International Workshop on Charm Physics (CHARM 2015)\\
Detroit, MI, 18-22 May, 2015
\end{Presented}
\vfill
\end{titlepage}
\def\thefootnote{\fnsymbol{footnote}}
\setcounter{footnote}{0}
%

\section{Introduction}
 CP violation in neutral meson-antimeson systems arises from three contributions: direct CPV, which is CP violation in a specific decay, and indirect CPV from mixing and from interference of decays with and without mixing. \\
  Mixing in decays of $D^0$ to CP eigenstates ($K^+K^-, \pi^+\pi^-$)  results in a different effective lifetime than that of decays to flavour eigenstates ($K^-\pi^+$). The observable $y_{CP}$ is defined as
 \begin{equation}
  y_{CP}=\frac{\tau(D^0 \to K^-\pi^+)}{\tau(D^0 \to K^+K^-)} -1 \quad.
 \end{equation}
 CP violation gives rise to a difference of effective lifetimes of $D^0$ and $\overline{D}^0$ decays to the same CP eigenstate. The asymmetry $A_{\Gamma}$ can be defined as
\begin{equation}
A_{\Gamma} = \frac{\tau(\overline{D}^0 \to K^-K^+) - \tau(D^0 \to K^+K^-)}{\tau(\overline{D}^0 \to K^-K^+) + \tau(D^0 \to K^+K^-)}  \quad.
\end{equation}
It follows that if no indirect CP violation is present,  $A_{\Gamma} $ becomes zero.\\
The Standard Model prediction for CP violation in charm is of the order $10^{-3}$ \cite{Grossman:2006jg}. Any measurements of a larger value would indicate signs of New Physics, making this field an interesting area of study.

\section{Time-dependent analysis of $D^0 \to K^+K^-, \pi^+\pi^-$}

Through a measurement of  $y_{CP}$ and $A_{\Gamma}$ we measure indirect CP violation, which is common to all $D^0$ decay modes. The quantities $y_{CP}$ and $A_{\Gamma}$ are extracted via proper decay time measurement. We present here the final result of the update of the same analysis, performed by the Belle Collaboration in 2007~\cite{old-ygamma}. The new results cover the entire Belle data sample, corresponding to 976 fb$^{-1}$ collected at or near $\Upsilon$(4S), $\Upsilon$(1S), $\Upsilon$(2S), $\Upsilon$(3S) and $\Upsilon$(5S) resonances. An improved analysis method is implemented, which recognises two different configurations of the Silicon Vertex Detector (SVD1 and SVD2) that were used during the data taking~\cite{SVD2}, and accounts for the polar angle dependence.

\subsection{Event Selection}
The $D^0$ is required to come from a decay chain $D^{*+} \to D^0 \pi^+_{s}$. The charge of the slow pion provides the necessary tag on the flavour of the charm meson. Additionally, it enables us to set a constraint on the total energy released in the decay, $q=m(D^{*+}) - m(D^0) - m(\pi^+_{s})$, which provides excellent background suppression.\\
To select kaons and pions, standard Belle particle identification is applied~\cite{PID}. The daughters of the $D^0$ are refitted to a common vertex. For the vertex fit of the $D^{*+}$,  $D^0$ and $\pi^+_{s}$ are also fitted to the interaction point. In both cases, a confidence level greater that $10^{-3}$ is required. To exclude $D^0$s coming from decays of $B$ mesons, a cut on the CMS momentum of the $D^{*+}$ is imposed. It is required that $p_{CMS}(D^{*+})>$ 2.5 GeV (3.1 GeV for $\Upsilon$(5S)). Candidates for $D^0$s are selected based on the $D^0$ mass $m(D^0$) and $q$, with a window imposed around the nominal value. For $m(D^0$), the allowed deviation is $\pm$2.25 $\sigma_{M}$, where $\sigma_{M}$ is the R.M.S. width of the $D^0$ peak ($\sigma_M$) and depends on  the decay mode and SVD configuration. Typical values are $\sigma_{M} \approx 6-8$ MeV.  The window in $q$ is $\pm$0.66 MeV (0.82 MeV) for SVD1 (SVD2). The selection criteria are optimised in order to achieve the minimal statistical error on $y_{CP}$. The proper decay time of the $D^0$ is calculated from the projection of the vector joining the two vertices $\vec{L}$ onto the $D^0$ momentum vector: $t= m_{nom}(D^0) \vec{L} \cdot \vec{p} / p^2 $, where $m_{nom}(D^0)$ is the nominal $D^0$ mass. The proper decay time uncertainty is evaluated from the error matrices of the production and decay vertices. It is required $\sigma_t < 440$ fs  (370 fs) for SVD1 (SVD2) to reject candidates with poorly determined proper decay time. 

\begin{figure}
\subfigure[]{
 \includegraphics[width=0.3\textwidth]{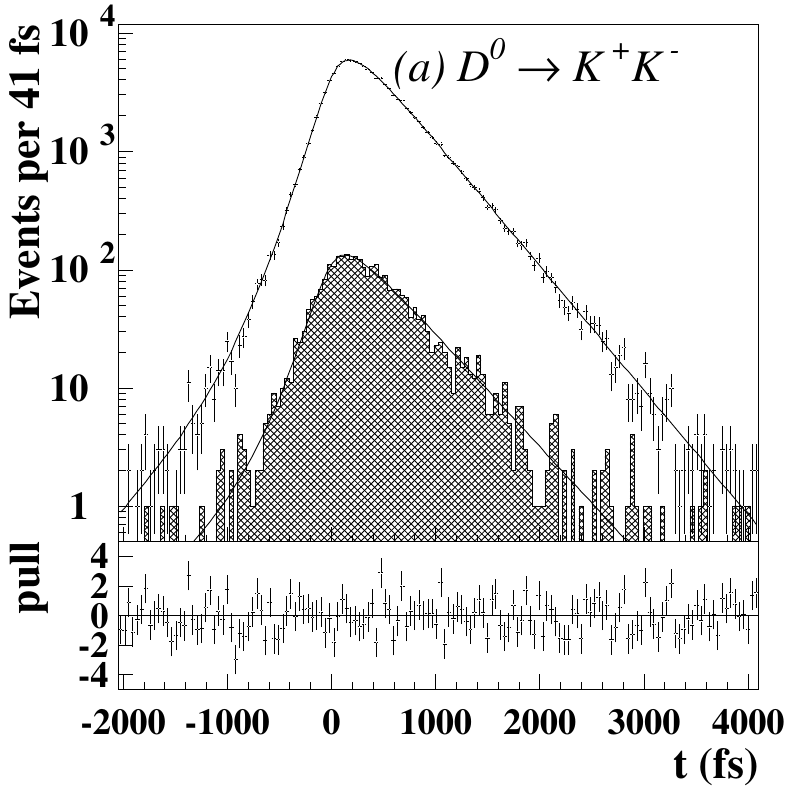}
}
\subfigure[]{
\includegraphics[width=0.3\textwidth]{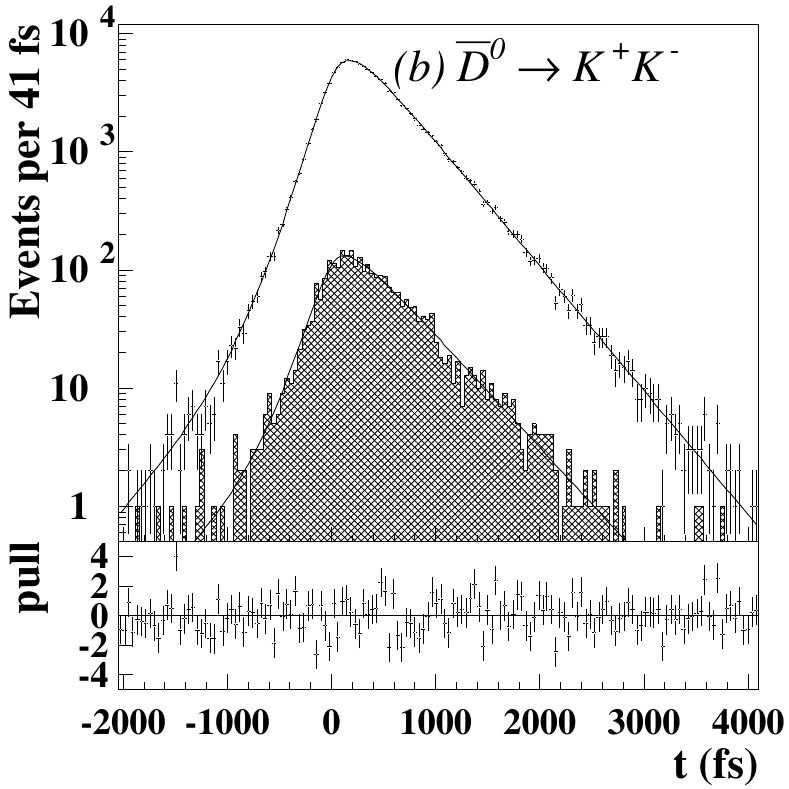}
}
\subfigure[]{
 \includegraphics[width=0.3\textwidth]{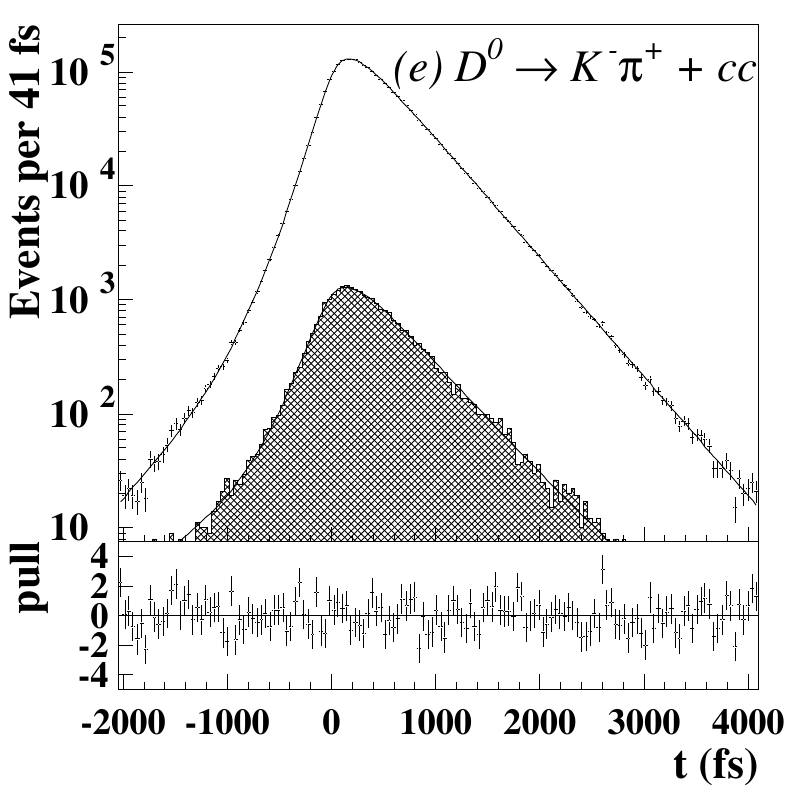}\
}

\subfigure[]{
\includegraphics[width=0.3\textwidth]{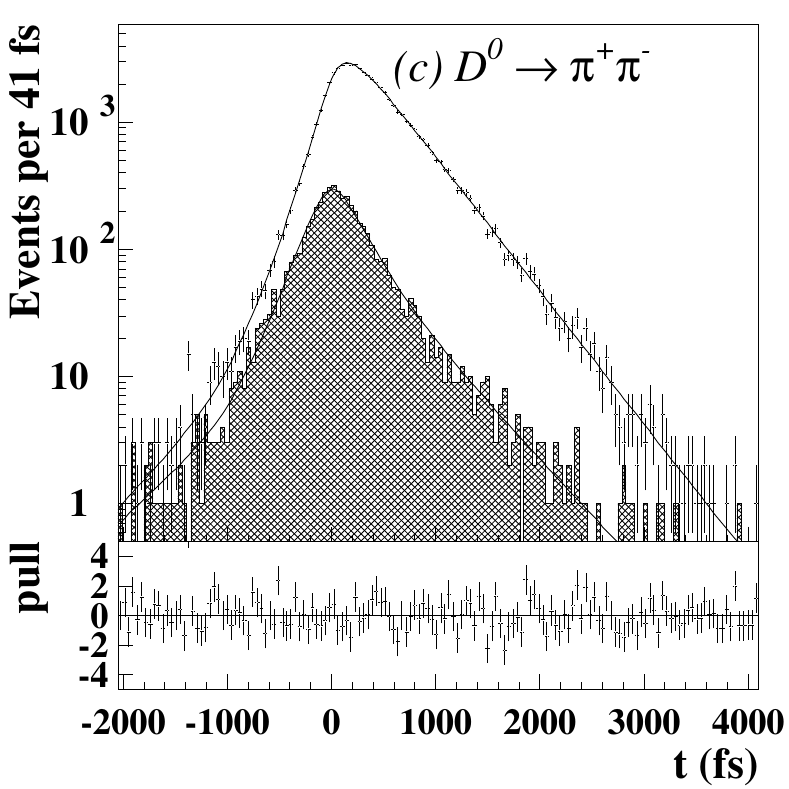}
}
\subfigure[]{
 \includegraphics[width=0.3\textwidth]{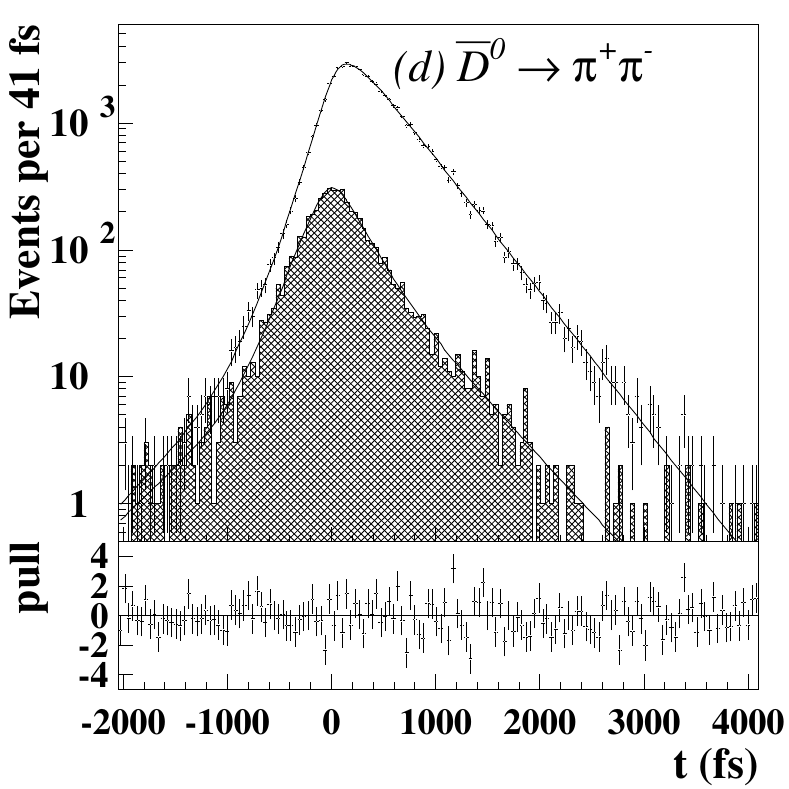}\
}
\caption{Proper decay time distributions summed over cos($\theta^*$) bins and both SVD configurations with the sum of fitted functions superimposed. Shown as error bars  are  the  distributions  of  events  in  the $m(D^0)$
signal  region  while  the  shaded area represents background contributions as obtained from sidebands. The corresponding pull is shown below each plot.}
\label{fig:lifetime-fit}
\end{figure}

 The proper decay time distribution is parameterised as
 \begin{equation}
F(t) = \frac{N}{\tau} \int e^{-t'/\tau}R(t-t')dt' +B(t) \quad,
\end{equation}
where $\tau$ is the effective lifetime, $N$ is the signal yield, $R(t)$ is the resolution function and $B(t)$ is the background distribution. Background is fixed from a fit to the sideband distribution. The position of the sidebands is optimised as to minimise the systematic uncertainty. The resolution function is  constructed using a normalised distribution of $\sigma_t$, combining fits for different bins. An additional offset parameter is introduced to correct for a certain misalignment of the SVD detector. As this  parameter is a function of the cosine of the $D^0$ CMS polar angle $\theta^*$, the resolution function is evaluated in separate bins of cos($\theta^*$). A simultaneous binned maximum likelihood fit is performed in each bin for all three channels, separately for SVD1 and SVD2. The fit results are shown in Figure~\ref{fig:lifetime-fit}. \\
Obtaining measurements of $y_{CP}$ and $A_{\Gamma}$ for all bins, the results are combined with least squares fit to constant to obtain the final result.\\
The fit is tested on Monte Carlo simulation equivalent to six times the data statistics. A linearity test shows that no bias is present.

\subsection{Systematics and Final Result}
The estimated systematic uncertainties are listed in Table~\ref{tab:syst-uncertainties}. The main contribution arises from the misalignment of the SVD and is estimated using Monte Carlo simulations for different misalignments. The uncertainty due to the position of the mass window is estimated by varying the position of the window. The uncertainty on background comprises a contribution due to statistical fluctuations and a contribution that arises from modelling the distribution, which is estimated from MC simulation. The two contributions are added in quadrature. Systematics due to the resolution function are estimated using  alternative parameterisations. Systematics due to binning are estimated by varying the number  of  bins in cos($\theta^*$) and $t$. All individual systematic uncertainties are added in quadrature to obtain the overall uncertainty.\\
\begin{table}
\centering
  \begin{tabular}{c |c c}
  Source & $\Delta$y$_{CP}$ (\%) &  $\Delta$A$_{\gamma}$ (\%)\\
  \hline
\hline 
  SVD misalignment &  0.060 & 0.041 \\
  Mass window position & 0.007 & 0.009\\
  Background & 0.059 & 0.050\\
  Resolution function & 0.030 & 0.002\\
  Binning & 0.021 & 0.066\\
\hline
Total & 0.092 & 0.066\\
\hline
\hline
  \end{tabular}
\caption{Systematic uncertainties for the time-dependent analysis of $D^0 \to  K^+K^-, K^-\pi^+, \pi^+\pi^-$.}
  \label{tab:syst-uncertainties}
 \end{table}

The final result for the difference in the effective lifetime of $D^0$ mesons decaying to CP-even eigenstates $K^+K^-, \pi^+\pi^-$ and the flavour eigenstate $K^-\pi^+$ is 
\begin{equation}
\nonumber
y_{CP} =[1.11 \pm 0.22(\mathrm{stat.}) \pm 0.09(\mathrm{syst.})]\% \quad,
\end{equation}
which corresponds to a 4.7 $\sigma$ significance when statistical and systematic error are added in quadrature. Simultaneously, the CP asymmetry was measured, obtaining
\begin{equation}
A_{\Gamma} = [-0.03 \pm 0.20(\mathrm{stat.}) \pm 0.07 (\mathrm{syst.})]\% \quad,
\end{equation}
which is consistent with no CP violation.

\section{Time-integrated analysis of $D^0 \to K^+K^-, \pi^+\pi^-$}

While the time-dependent analysis only measures indirect CP violation, it is possible to measure both direct and indirect CPV performing a time-integrated analysis of decay rates of neutral charm mesons to CP-even final states $K^+K^-$ and $\pi^+\pi^-$.  The presented results are an update of a 2008 analysis by the Belle Collaboration~\cite{old-acp} using the full Belle data sample. The asymmetry of the time-integrated decay rates of $D^0 \to f$ and $\overline{D}^0 \to \overline{f}$ is defined as
\begin{equation}
 A_{CP} = \frac{\Gamma(D^0 \to f) - \Gamma(\overline{D}^0 \to \overline{f})}{ \Gamma(D^0 \to f) +\Gamma(\overline{D}^0 \to \overline{f})} \quad,
 \end{equation}
 where $\Gamma$ is the partial decay width. The flavour of the charm meson is tagged via the charge of the slow pion from the decay  $D^{*+} \to D^0 \pi^+_{s}$. However, the experimentally measured quantity is 
 \begin{equation}
  A_{raw} = \frac{N(D^0 \to f) - N(\overline{D}^0 \to \overline{f})}{N(D^0 \to f) + N(\overline{D}^0 \to \overline{f})} \quad,
 \end{equation}
 where $N$ is the number of events from a certain decay. This quantity, called raw asymmetry, comprises besides the physical CP asymmetry also the production and detector-induced asymmetry:   $A_{raw} = A_{CP} + A_{FB} + A_{\varepsilon^{\pm}}$. The production asymmetry $A_{FB} $ is a forward-backward asymmetry in production of $D^{*+}$ and  $D^{*-}$ and arises from $\gamma - Z^0$ interference and higher order QED effects in $e^+e^- \to c\overline{c}$. It is assumed to be the same for all charm mesons. The term $A_{\varepsilon^{\pm}}$ is a detector-induced asymmetry that arises from  different reconstruction efficiencies for  positively and negatively charged particles. Since the $D^0$ decays in question are self-conjugated, the only charged particle that contributes to the asymmetry $A_{\varepsilon^{\pm}}$ is the slow pion, with $A_{\varepsilon^{\pm}}$ hence becoming $A_{\varepsilon}(\pi_{s}^\pm)$. This term is determined using tagged and non-tagged decays $D^0 \to K^- \pi^+$, where it holds
 \begin{align}
 A_{untag}&= A_{FB}(D^0)  + A_{CP}(K\pi) + A_{\varepsilon}(K\pi) \quad,\\
 A_{tag}&= A_{FB}(D^{*+})  + A_{CP}(K\pi) + A_{\varepsilon}(K\pi)  + A_{\varepsilon}(\pi_{s}^\pm) \quad.
 \end{align}
 Since the forward-backward asymmetry is assumed to be the same for all charm mesons, it follows that
 $A_{\varepsilon}(\pi_{s}^\pm) = A_{tag} - A_{untag}$. Because $A_{\varepsilon}(K\pi)$ and $A_{\varepsilon}(\pi_{s}^\pm)$ are functions of the corresponding phase spaces in the laboratory frame, $A_{\varepsilon}(\pi_{s}^\pm) $ is corrected for separately in bins of $p_{\pi_{S}}$ and $\theta_{\pi_{S}}$.\\
 The forward-backward asymmetry is odd function of $\theta^*$ and is corrected for using
 \begin{align}
 A_{CP} &= \frac{1}{2} [ A_{raw}^{corr}(\cos\theta^*) + A_{raw}^{corr}(-\cos\theta^*)] \quad,\\
 A_{FB} &= \frac{1}{2} [ A_{raw}^{corr}(\cos\theta^*) - A_{raw}^{corr}(-\cos\theta^*)] \quad,
 \end{align}
 where $A_{raw}^{corr}$ is the raw asymmetry after $A_{\varepsilon}(\pi_{s}^\pm)$ correction.
 
 \subsection{Event Selection}
\begin{figure}
 \centering
  \includegraphics[width=0.7\textwidth]{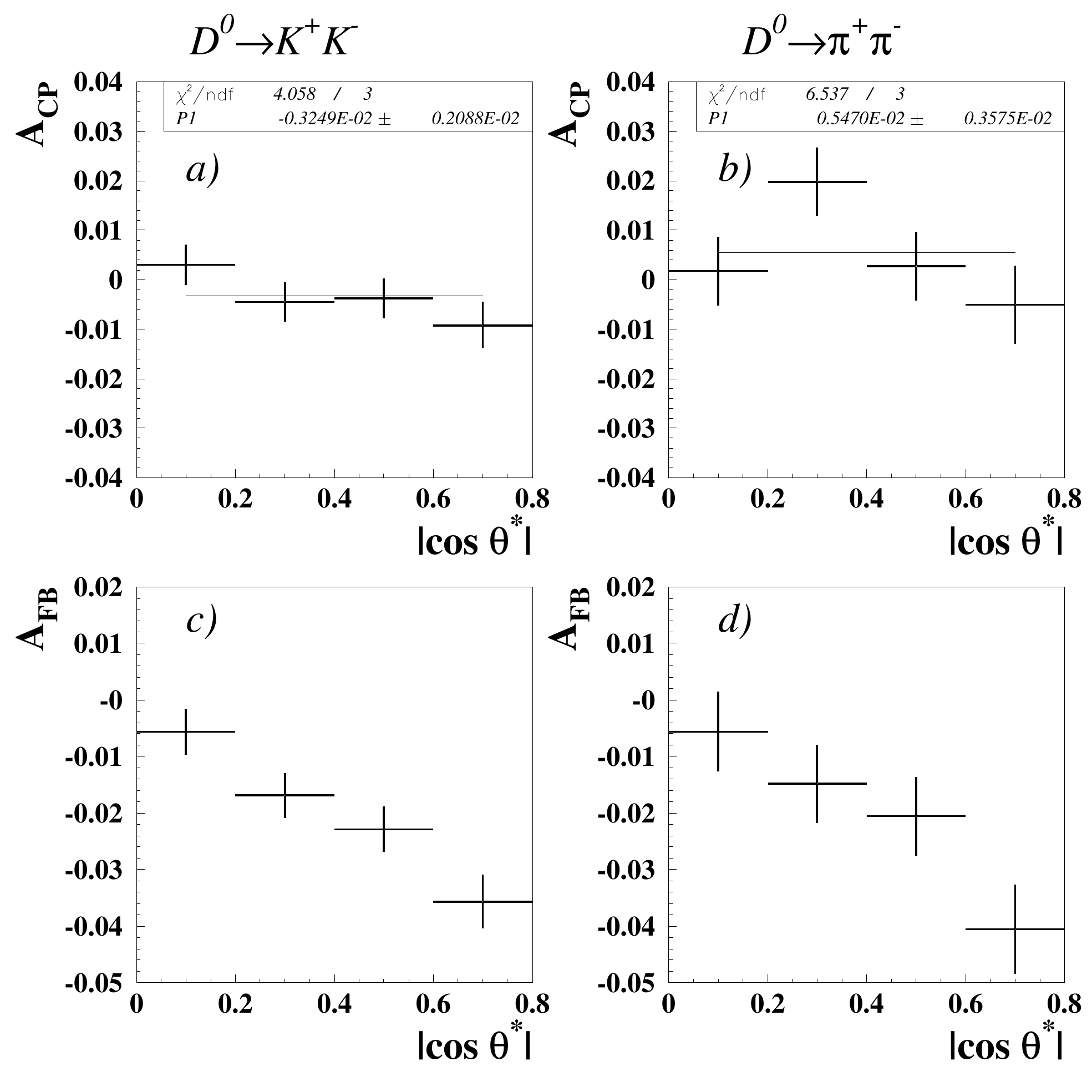}
  \caption{CP-violating asymmetries in $K^+K^-$ and $\pi^+\pi^-$ final states, and forward-backward asymmetries for the same two final states.}
    \label{fig:ACP-AFB}
 \end{figure}

 The vertex fit and p$_{CMS}$(D$^{*+}$) cuts are the same as stated in the previous Section. Other selection criteria are optimised so that the error on the asymmetry is minimal. The windows in $m(D^0)$ ($q$) are  $\pm$17.8 (1.00) MeV for $K^+K^-$, $\pm$17.8 (1.85) MeV for  $K^-\pi^+$ and $\pm$17.2 (0.90) MeV for $\pi^+\pi^-$.\\
 The signal yield is extracted via background subtraction. Background in the signal window is estimated based on two symmetrical sidebands that together amount to the width of the signal window, with their position starting at $\pm 20$ MeV from the nominal $D^0$ mass. Additionally, the range is limited to $|$cos($\theta^*)|<$ 0.8 in order to decrease $A_{\varepsilon}(\pi_{s}^\pm)$-related systematics.\\
 The physical asymmetry $A_{CP}$ is extracted in bins of cos($\theta^*$), with the final result obtained via a fit to constant. The corresponding plots are shown in Figure~\ref{fig:ACP-AFB}.
 
 \subsection{Systematics and Final Results}
 The estimated systematic uncertainties are listed in Table~\ref{tab:syst-uncertainties2}. The uncertainty corresponding to the signal counting method comprises two contributions, summed in quadrature: possible small differences in the signal shapes of $D^0$ and $\overline{D}^0$ (uncertainty estimated from the tagged $K^-\pi^+$ sample) and in background between the signal window and sidebands. The uncertainty corresponding to sideband selection is estimated by varying the position of the sidebands. The uncertainty of the $A_{\varepsilon}(\pi_{s}^\pm)$ correction arises from statistics of the $K^-\pi^+$ sample and binning (estimated from using different binnings and requiring different minimal statistics per bin). The extraction method of $A_{CP}$ contributes a systematic uncertainty due to binning and two different configurations of SVD.\\
 \begin{table}
 \centering
  \begin{tabular}{c |c | c| c }
  & $A_{CP}^{K^+K^-} (\%)$ & $A_{CP}^{\pi^+\pi^-} (\%) $& $\Delta A_{CP}$ (\%)\\
  \hline
\hline 
  Signal counting method & 0.055 & 0.023 & 0.037\\
  $\pi_{S}$ correction & 0.065 & 0.067 & 0.014\\
  $A_{CP}$ extraction method & 0.06 & 0.050 & 0.051\\
\hline
Total & 0.085 & 0.087 & 0.064\\
\hline
\hline
  \end{tabular}
\caption{Systematic uncertainties for the time-integrated analysis of $D^0 \to  K^+K^-, \pi^+\pi^-$.}
 \label{tab:syst-uncertainties2}
\end{table}
 The final results for the CP asymmetries are
 \begin{align}
 A_{CP}^{KK} &= [-0.32 \pm 0.21(\mathrm{stat.}) \pm 0.09 (\mathrm{syst.})]\% \quad, \\
 A_{CP}^{\pi\pi} &= [\phantom{-}0.55 \pm 0.36(\mathrm{stat.}) \pm 0.09 (\mathrm{syst.})]\%\quad, \\
\Delta A_{CP} &= [-0.87 \pm 0.41(\mathrm{stat.}) \pm 0.06 (\mathrm{syst.})]\% \quad,
 \end{align}
and are consistent with no CPV.


 \section{Time-integrated analysis of $D^0 \to \pi^0\pi^0$}
 We report also a time-integrated analysis of CP violation in decays $D^0 \to \pi^0\pi^0$~\cite{nisar}, performed on 966 fb$^{-1}$ of Belle data. The analysis is largely similar to the time-integrated analysis  described in the previous Section. 
  \subsection{Event Selection}
\begin{figure}[H]
\centering
\subfigure[]{
\includegraphics[width=0.32\textwidth]{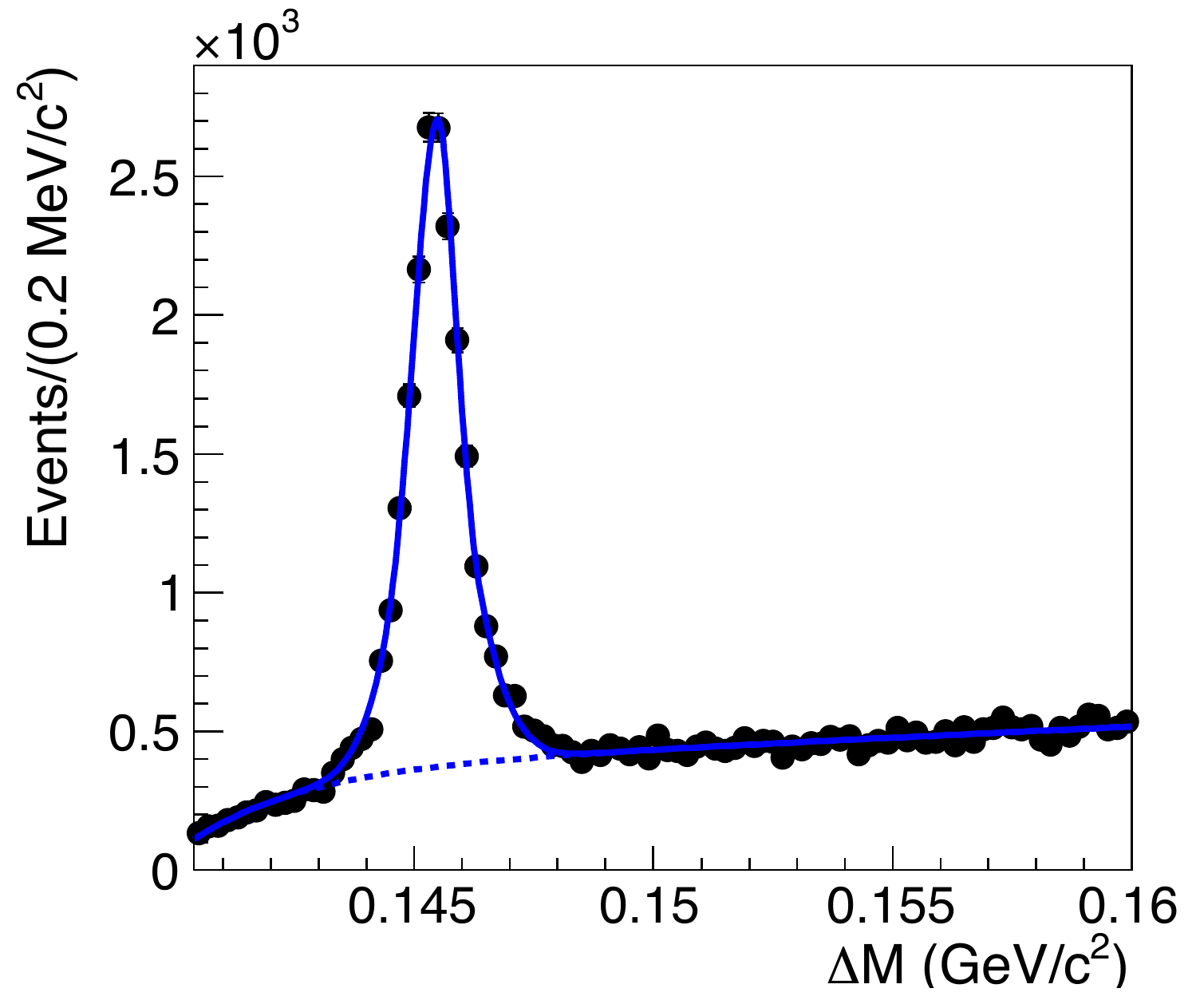}
}
\subfigure[]{
\includegraphics[width=0.32\textwidth]{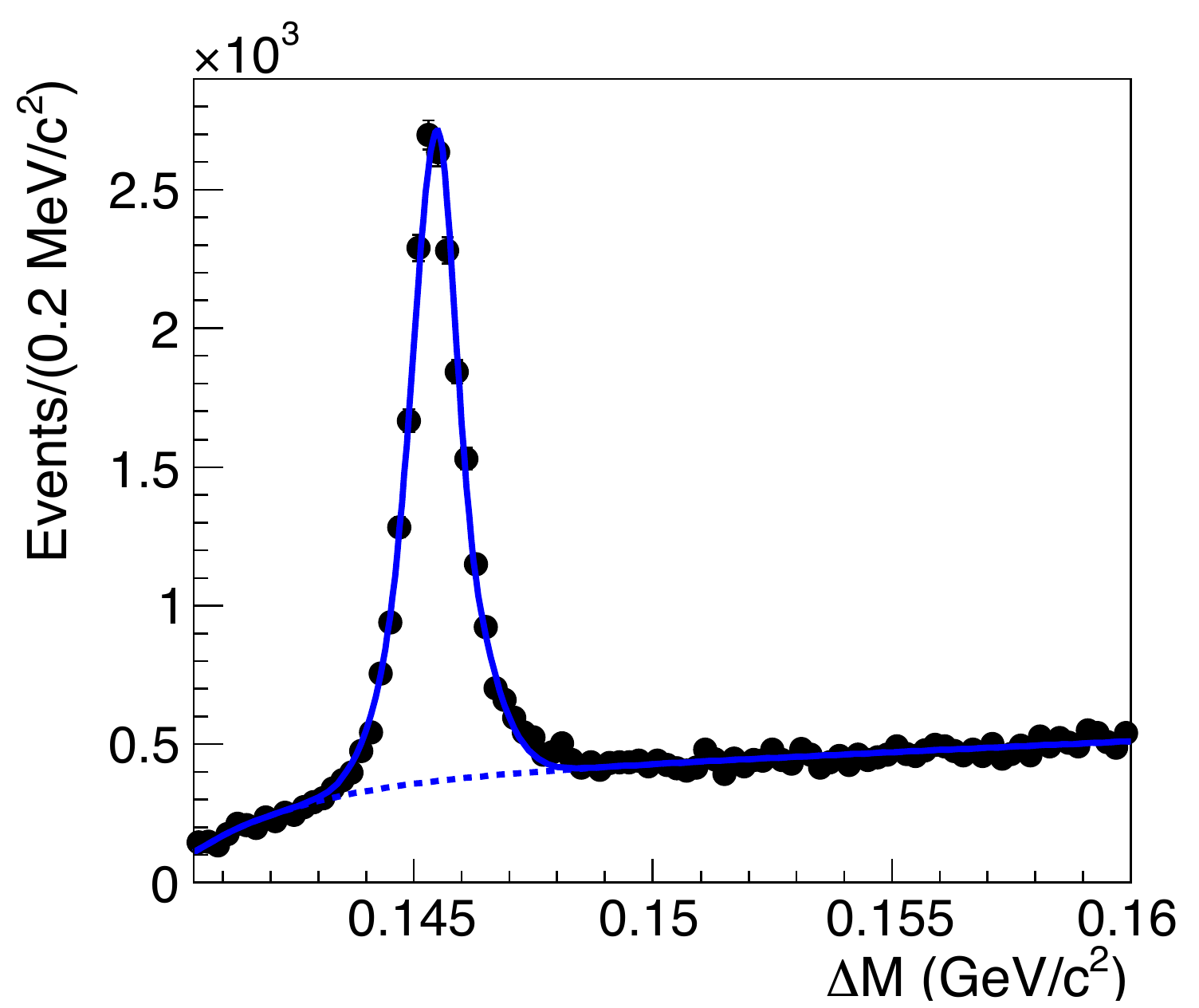}
}
\subfigure[]{
\includegraphics[width=0.275\textwidth]{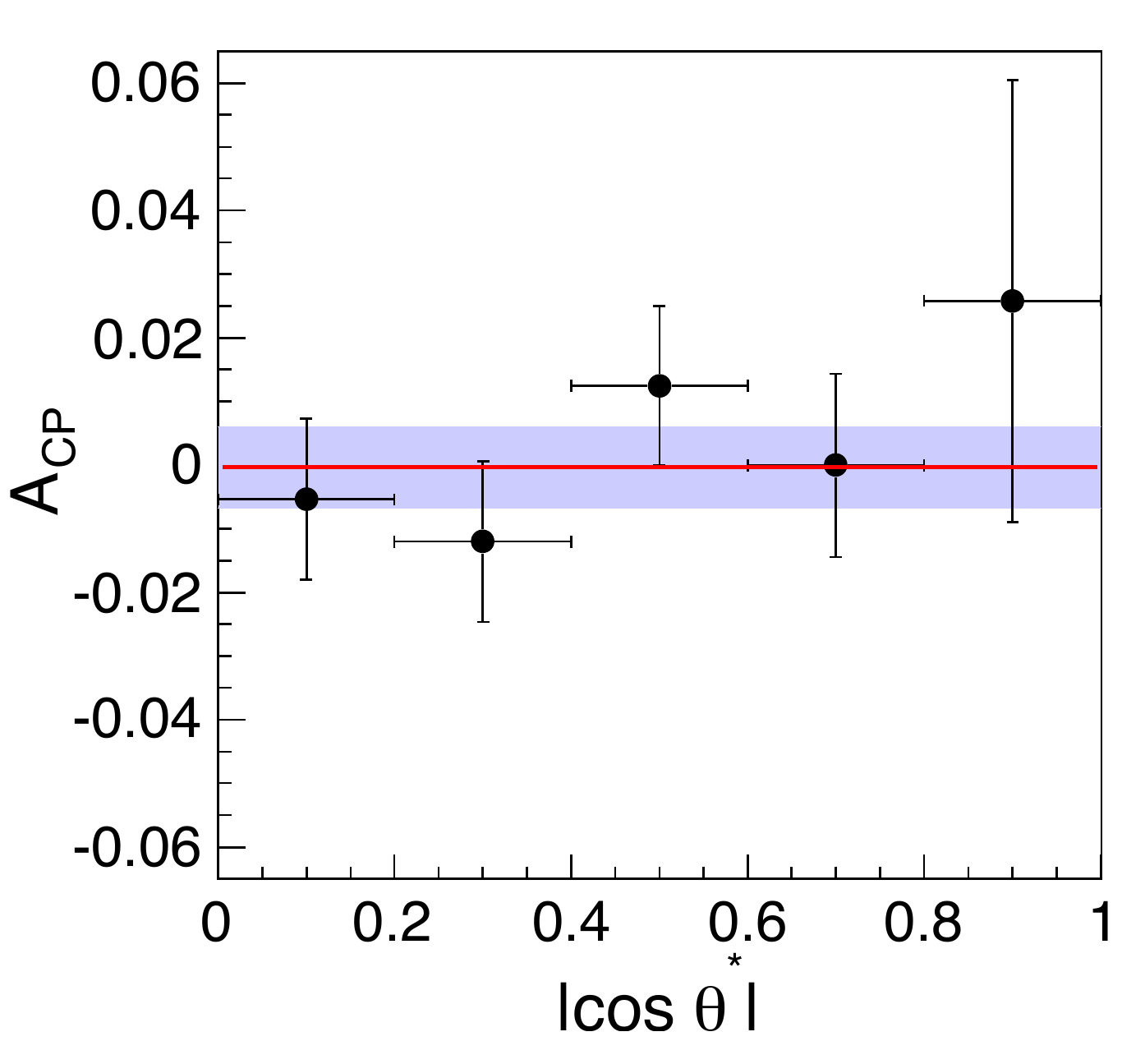}
}
\caption{Projection of the simultaneous fit in $\Delta m$ to the $D^0$ (left) and $\overline{D}^0$ (middle), and $A_{CP}$ in bins of cos$(\theta^*)$ (right).}
\label{fig:dm-fit}
\end{figure}

The $D^0$ is required to originate from the decay $D^{*+} \to D^0 \pi^+_{s}$. The same p$_{CMS}$(D$^{*+}$) cut as before is applied. The optimisation of selection criteria is done in order to minimise the error on the raw asymmetry. The  obtained range of $m(D^0)$ is (1.758, 1.930) GeV and the $q$ range is (0.14, 0.16) GeV. \\
The corrections for production and detector-induced asymmetry is performed as previously described. The signal yield is extracted via a simultaneous fit in $\Delta m$  of  $D^0$ and $\overline{D}^0$ samples. The fit is performed in bins of (cos$(\theta^*), p_{T}^{\pi_{S}}, \cos(\theta^{\pi_{S}}))$, and the final result obtained via a fit to constant on values in bins of $\cos(\theta^*)$. The procedure is tested and confirmed on Monte Carlo simulations. The corresponding plots are shown in Figure~\ref{fig:dm-fit}.

\subsection{Systematics and Final Result}
The estimated systematic uncertainties are listed in Table~\ref{tab:syst-uncertainties3}. The final result for the CP asymmetry in decays $D^0 \to  \pi^0\pi^0$ is
\begin{equation}
A_{CP} = [-0.03 \pm 0.64(\mathrm{stat.}) \pm 0.10 (\mathrm{syst.})]\% \quad,
\end{equation}
which is consistent with no CP violation. It is the most accurate measurement to date.
 \begin{table}[H]
 \centering
  \begin{tabular}{c |c }
  \hline
\hline 
  signal shape & 0.03\\
  $\pi_{S}$ correction & 0.07\\
  $A_{CP}$ extraction method & 0.07\\
\hline
Total & 0.10\\
\hline
\hline
  \end{tabular}
\caption{Systematic uncertainties for the time-integrated analysis of $D^0 \to  \pi^0\pi^0$.}
  \label{tab:syst-uncertainties3}
\end{table}

\Acknowledgements

We thank the KEKB group and all institutes and agencies that support the work of the members of the Belle Collaboration.

\end{document}